\documentclass{ifacconf}

\usepackage{graphicx}      
\usepackage{natbib}       
\usepackage{myMacros}
\usepackage{lipsum}
\usepackage{mathtools}
\usepackage{cuted}
\usepackage{url}
\usepackage{booktabs}
\usepackage{tikz}
\usepackage[locale=US]{siunitx}
\DeclareSIUnit{\degree}{deg}
\DeclareSIUnit{\bar}{bar}

\newcommand\copyrighttext{\textcopyright 2023 the authors. This work has been accepted to IFAC for publication under a Creative Commons Licence CC-BY-NC-ND.}
\newcommand\copyrightnotice{
\begin{tikzpicture}[remember picture,overlay]
    \node[anchor=north,yshift=-10pt] at (current page.north) {\shortstack{Accepted for publication at the IFAC World Congress 2023\\ \\
		  \fbox{\parbox{\dimexpr\textwidth-\fboxsep-\fboxrule\relax}{\copyrighttext}}} };
\end{tikzpicture}%
}

\begin{document}

\copyrightnotice

\begin{frontmatter}

\title{Automated Tuning of Nonlinear Kalman Filters for Optimal Trajectory Tracking Performance of AUVs\thanksref{footnoteinfo}}

\thanks[footnoteinfo]{This publication results from the project line "Technologies for Rapid Ice Penetration and Subglacial Lake Exploration" (TRIPLE: \url{https://triple-project.net/the-nanoauv/}) and is supported by the German Federal Ministry for Economic Affairs and Energy (grant 50NA2009).}

\author[First]{Maximilian Nitsch} 
\author[First]{David Stenger} 
\author[First]{Dirk Abel}

\address[First]{Institute of Automatic Control, RWTH Aachen University, Aachen, Germany (e-mail: m.nitsch@irt.rwth-aachen.de).}

\begin{abstract}    
The performance of navigation algorithms significantly determines the trajectory tracking accuracy of the guidance, navigation, and control (GNC) system of an autonomous underwater vehicle (AUV). 
In closed-loop operation, the interaction among path planning, control, and navigation
plays a crucial role in the tracking accuracy of the overall GNC system. 
A Doppler velocity log (DVL) is often used for AUVs to measure velocity over the ground, positively affecting the closed-loop tracking error. 
However, a DVL may not be installed in miniaturized AUVs due to limited space and energy.
In this paper, a navigation filter for an underactuated miniature AUV (nanoAUV) is considered that is mainly based on acoustic localization using a novel highly-miniaturized ultra-short baseline (USBL) system and a depth pressure sensor. 
The nanoAUV is being developed for subglacial lake exploration.
We compare two unscented Kalman filters (UKF) with different prediction models - the classical strapdown inertial navigation systems (SINS) model and a hydrodynamic motion model (HMM).
To enable a fair comparison, filter parameters are auto-tuned with Bayesian optimization (BO) for open- and closed-loop performance, which is novel in AUV navigation. The results indicate that BO performs similarly to particle swarm optimization (PSO) regarding sample efficiency for the proposed problem.
To quantify the GNC tracking performance, we use extensive Monte Carlo simulations.
Results suggest that with BO-tuned navigation filter parameters, the median tracking error is reduced by up to 50\% compared to default parametrization. 
\end{abstract}


\end{frontmatter}

\section{Introduction}
To explore unknown waters with harsh environmental conditions such as under ice, a \textbf{GNC system} of an AUV must achieve high trajectory tracking accuracy and operate robustly since surfacing is not possible in case of an emergency.
Such under-ice explorations are the task of the nanoAUV, which is currently being developed within the \textbf{TRIPLE project line}. 
After a successful demonstration, the nanoAUV will explore subglacial waters in Antarctica. Its first mission is planned at Neumayer III Station in Antarctica under the Ekström Ice Shelf. The task includes autonomous docking procedures and exploration of the water column.

The navigation of most AUVs relies on sensor fusion of an inertial measurement unit (IMU), a Doppler velocity log (DVL), a depth pressure sensor, and a magnetometer (MAG). Many AUVs make use of terrain-aided navigation techniques or acoustic positioning systems as long baseline (LBL) or (primarily heavy and bulky) ultra-short baseline (USBL) systems, according to \cite{hagen2007}.
However, due to limited installation space and energy resources, terrain-aided navigation or DVLs may not be suitable for small AUVs like the nanoAUV. Instead of the DVL, only a USBL is used because it allows sending and receiving telemetry data and ensures a complete drift correction of the SINS. This poses a significant challenge for navigation because positioning mainly relies on the USBL performance, which strongly depends on the conditions of the acoustic channel. 

One way to address this issue is to use \textbf{HMM prediction} similar to \cite{hegrenaes2008} to mitigate USBL losses and the absence of a DVL. 
It is compared to the classical \textbf{SINS prediction} described by \cite{titterton2004}. Therefore, two unscented Kalman filters (UKF) with these prediction models are presented in this work.

\begin{figure}[htbp]
    \centering
    \includegraphics[width=1.0\columnwidth]{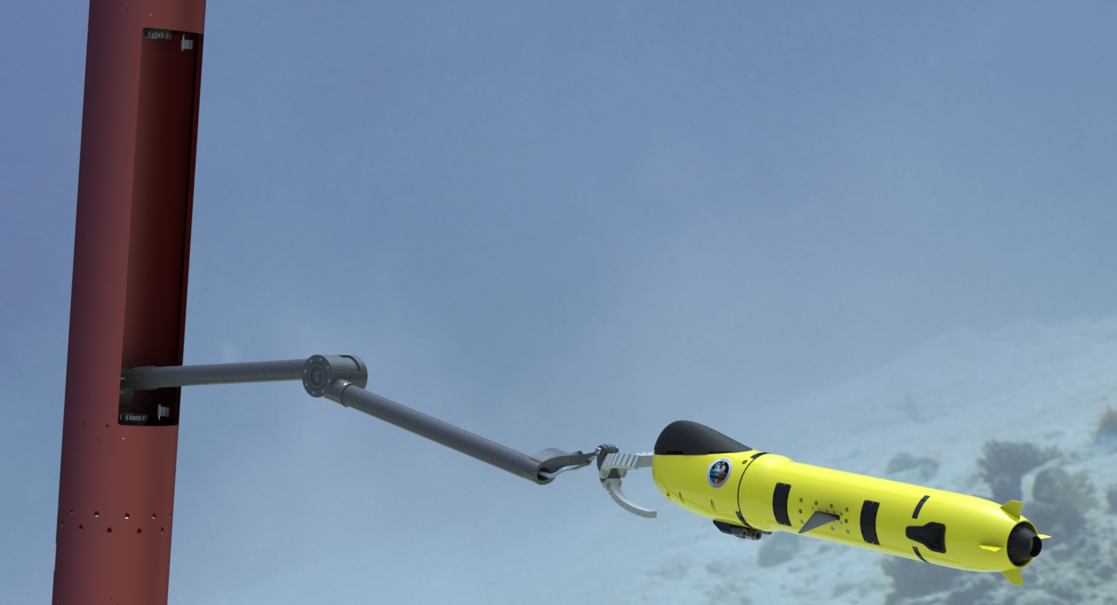}
    \caption{The nanoAUV docking with the Launch- and Recovery System (Credits: Sebastian Meckel, MARUM).}
    \label{fig:nanoauv_docking}
\end{figure}

An automatic tuning approach for the covariance matrices based on black-box optimization is used to fairly and systematically compare the filter configurations. The navigation performance is evaluated on several representative trajectories, and \textbf{Bayesian optimization (BO)} is used to search for best-performing filter parameters. This ensures that differences in navigation performance can be attributed to the filter structure and not the parametrization. \cite{LeRiche.2021} suggested BO be very sample efficient, especially for small budgets and low dimensional problems. In addition to tuning the open-loop (OL) state estimation accuracy, the filter is being tuned for \textbf{optimal tracking performance} in closed-loop (CL) operation with the complete GNC system. BO has been used for filter parameter optimization,e.g., in \cite{Gehrt.1123202011242020}.
However, except for our previous work \cite{Stenger.2022}, automatic tuning for filter parameters with BO has not been considered yet for AUVs and is still mostly done by hand by the engineer, according to \cite[p.\,22]{carpenter2018}. 
\cite{levi2019} tuned filter parameters of an AUV with PSO. Although PSO has been shown to be less sample efficient than BO cf., e.g., \cite{Stenger.2022b}, in this work, we compare BO and PSO in terms of sample efficiency. 
In our previous work \cite{Stenger.2022}, we focused on the energy efficiency of the overall GNC system. Here, the focus is on the performance comparison of the HMM and SINS filter structures in the open- and closed-loop operations. 

Given a representative trajectory, the two filter architectures' open- and closed-loop performance is validated with extensive \textbf{Monte Carlo simulations} for multiple sensor noise realization, water currents, and mismatch of the nominal HMM and USBL outages. The results indicate that optimized parameters for state estimation accuracy do not lead to optimal tracking performance in closed-loop operation. It is also shown that using the SINS prediction for closed-loop operation is more advantageous than an HMM-based prediction under model mismatch.

This paper is structured as follows: first, in Sec.\,\ref{sec:simulation_environment}, the nanoAUV model and the sensor simulation environment are described. The navigation filters and parametrization procedures are shown in Sec.\,\ref{sec:navigation_filter}. The results are given in Sec.\,\ref{sec:results} followed by the conclusions in Sec.\,\ref{sec:conclusions}.

\section{Simulation Environment}\label{sec:simulation_environment}

\subsection{Hydrodynamic Motion Model}
The well-known HMM, according to \cite{fossen2006}, is used to model the six degrees of freedom of the nanoAUV:
\begin{align} \label{eq:fossen_model}
    \myFrameVec{M}{RB}{} \myFrameVecDot{\nu}{}{} &+ \myFrameVec{C}{RB}{}(\myFrameVec{\nu}{}{})\myFrameVec{\nu}{}{} + 
    \myFrameVec{M}{A}{} \myFrameVecDot{\nu}{r}{} + \myFrameVec{C}{A}{}(\myFrameVec{\nu}{r}{})\myFrameVec{\nu}{r}{} + \dots \\
    & \myFrameVec{D}{}{}(\myFrameVec{\nu}{r}{}) \myFrameVec{\nu}{r}{}+ \myFrameVec{g}{}{}(\myFrameVec{\eta}{}{}) = \myFrameVec{\tau}{}{} \nonumber \, \text{.}
\end{align} 
with mass matrices $\myFrameVec{M}{RB}{}$ and $\myFrameVec{M}{A}{}$, Coriolis-centripetal matrices $\myFrameVec{C}{RB}{}(\myFrameVec{\nu}{}{})$ and $\myFrameVec{C}{A}{}(\myFrameVec{\nu}{r}{})$, and damping matrix $\myFrameVec{D}{}{}(\myFrameVec{\nu}{r}{})$. Submerged weight and buoyancy forces are summed up in $\myFrameVec{g}{}{}(\myFrameVec{\nu}{}{})$. Because the exact hydrodynamics are not exactly known at this stage of development, we assume torpedo-shaped AUV and only linear damping $\myFrameVec{D}{}{}(\myFrameVec{\nu}{r}{})$.
The state vector $\myFrameVec{x}{}{} = [ \myFrameVec{\nu}{}{}, \myFrameVec{\eta}{}{} ]^\mathrm{T}$ consists of the motion components $\boldsymbol{\nu} = [ \myFrameVec{\nu}{1}{}, \myFrameVec{\nu}{2}{} ]^\mathrm{T} = [u,v,w,p,q,r]^\mathrm{T}$ in body frame and $\myFrameVec{\eta}{}{} = [ \myFrameVec{\eta}{1}{}, \myFrameVec{\eta}{2}{} ]^\mathrm{T} = [n,e,d,\phi, \theta, \psi]^\mathrm{T}$ denotes the positions $n,e,d$ of the body frame in the Earth fixed frame. The Euler angles are roll $\phi$, pitch $\theta$ and yaw $\psi$. 
The water current vector $\myFrameVec{v}{c}{n} = [\myFrameScalar{u}{c}{n}, \myFrameScalar{v}{c}{n}, \myFrameScalar{w}{c}{n}]^{\mathrm{T}}$ is considered to calculate the relative velocity $\myFrameVec{\nu}{r}{} = \myFrameVec{\nu}{1}{} - \myFrameVec{\nu}{c}{}$.   
The HMM delivers the ground-truth states for our sensor simulation models, which generate noisy measurements for the navigation filter. The generalized forces $\myFrameVec{\tau}{}{}$ serve as the input for the HMM. They are calculated as a function of the control input using idealized models for the buoyancy engine, movable mass, and three thrusters. It should be noted that the nanoAUV is underactuated. The sway velocity cannot be controlled independently of the yaw rate. 

\subsection{Water Current Simulation}
Similar to \cite{fossen2011}, we model the water current $\myFrameVec{v}{c}{}$ as first-order Gauss-Markov process:
\begin{align}\label{eq:water_current_simulation}
    \myFrameVecDot{v}{c}{n} = [\myFrameScalarDot{u}{c}{n}, \myFrameScalarDot{v}{c}{n}, \myFrameScalarDot{w}{c}{n}]^{\mathrm{T}} = -\mu \myFrameVec{V}{c}{n} + \myFrameVec{w}{V_c}{},
\end{align}
with inverse correlation time $\mu$ and white noise $\myFrameVec{w}{V_c}{}$. For this work, $\mu$ is set to $\SI{0.2}{\per\second}$. The standard deviations of $\myFrameVec{w}{V_c}{}$ are set to $[0.05, 0.05, 0.01]\,\SI{}{\metre\per\second}$. The solutions of \eqref{eq:water_current_simulation} are limited to $\pm [0.20, 0.20, 0.05]\,\SI{}{\metre\per\second}$.

\subsection{Sensor Noise Simulations}\label{subsec:sensor_simulation_models}
\subsubsection{Inertial Measurement Unit}
\begin{figure}[ht]
    \centering
    \includegraphics[width=1\columnwidth]{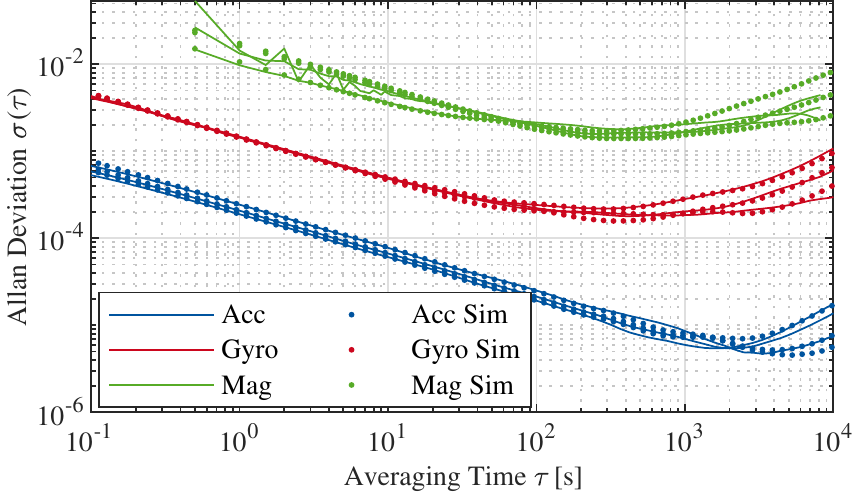}
    \caption{Real and simulated (dotted) Allan variances of accelerometer, gyroscope (STIM300 IMU) and magnetometer (RMI3100).}
    \label{fig:allan_variance}
\end{figure}

The simulation model of the IMU is based on a six-hour data log of a  tactical-grade Sensonor STIM300 IMU under static conditions. We identify the parameters of a first-order Gauss-Markov model according to \cite{farrell2022} with BO. 
The results are shown in Fig.\,\ref{fig:allan_variance}. The specific forces and rotation rates are simulated with \SI{100}{\hertz} according to:
\begin{align}\label{eq:imu_model}
\myFrameVecTilde{f}{ib}{b} =&~\myFrameVecDot{\nu}{1}{}- \myFrameVec{C}{n}{b}\myFrameVec{g}{nb}{n} + \myFrameVec{b}{on,a}{} + \myFrameVec{z}{N,a}{} + \myFrameVec{z}{B,a}{} + \myFrameVec{z}{K,a}{} \\
\myFrameVecTilde{\omega}{ib}{b} =&~\myFrameVec{\nu}{2}{} + \myFrameVec{b}{on,g}{} + \myFrameVec{z}{N,g}{} + \myFrameVec{z}{B,g}{} + \myFrameVec{z}{K,g}{},
\end{align}
with local gravity $\myFrameVec{g}{nb}{n} = [\SI{0}{},\SI{0}{},\SI{9.81}{}]^\mathrm{T}\SI{}{\metre\per\second}$, 
$\myFrameVec{b}{on}{}$ as the turn-on-bias and $\myFrameVec{z}{i}{}$ as the IMU stochastic error signals as in \cite{farrell2022}.
The values in \eqref{eq:imu_model} are fed to a quantization model afterward. Misalignment and temperature effects are neglected.

\subsubsection{Magnetometer}
A similar procedure as for the IMU is applied for the MAG simulation model. A six-hour (static) data log of a military-grade PNI RM3100 MAG is used to identify a stochastic error model similar to \eqref{eq:imu_model}:
\begin{align}\label{eq:magnetometer_simulation}
    \myFrameVecTilde{m}{nb}{b} = \myFrameVec{C}{n}{b} \myFrameVec{m}{nb}{n}+ \myFrameVec{z}{N}{} + \myFrameVec{z}{B}{} + \myFrameVec{z}{K}{},
\end{align}
with $\myFrameVec{m}{nb}{n}$ as the local magnetic field vector, which is determined from the world magnetic model for a given latitude and longitude. The data rate is \SI{100}{\hertz}.
The model \eqref{eq:magnetometer_simulation} neglects additional error sources like soft and hard iron disturbances and non-Gaussian errors due to magnetic anomalies. In real operations, the soft and hard iron disturbances are assumed to be calibrated before the mission starts. Magnetic anomalies are assumed to be suppressed by ensuring a proper EMC design of the AUV.

\subsubsection{Depth Sensor}
The nanoAUV depth is measured by means of an absolute pressure sensor. We use an RBRoem P sensor with a full-scale range of \SI{500}{\deci\bar} (\SI{500}{\metre} operating range). The sensor delivers the absolute water pressure with \SI{2}{\hertz}. The simulation model is:
\begin{align}\label{eq:depth_sensor_simulation}
    \myFrameScalarTilde{p}{abs}{} = \myFrameScalar{k}{p}{}\cdot d + \myFrameScalar{p}{0}{} + \myFrameScalar{w}{press}{},
\end{align}
with $\myFrameScalar{k}{p}{}=\SI{9806.38}{\pascal\per\metre H_2O}$, $\myFrameScalar{p}{0}{}=\SI{101325}{\pascal}$ as the atmospheric pressure and $\myFrameScalar{w}{press}{}$ as the white noise of the pressure measurements.
Furthermore, according to the datasheet, quantization, and saturation models are applied.

\subsubsection{Ultra-Short Baseline System}
A novel EvoLogics OEM USBL is used, which generates position fixes with $\SI{1}{\hertz}$. We simulate such measurements according to \cite{morgado2006}. 
The USBL is currently under development and miniaturized for the nanoAUV. The preliminary hydrophone positions $\myFrameVec{p}{HYDRO,i}{b}$ are used for this work. The developed USBL transceiver array consists of a modem and five hydrophones. The round-trip-time (RTT) between nanoAUV USBL transceiver and a (home) USBL transponder at $\myFrameVec{\eta}{USBL,H}{} = \myFrameVec{0}{}{}$ is modeled according to:
\begin{align}
    \myFrameScalarTilde{RTT}{}{} = \frac{2}{c}\cdot \| \myFrameVec{\eta}{USBL,AUV}{}-\myFrameVec{\eta}{USBL,H}{} \| + \myFrameScalar{w}{RTT}{},
\end{align}
with $c=\SI{1500}{\metre\per\second}$ and $\myFrameScalar{w}{RTT}{}$ as the white noise of the RTT measurement ($1\sigma = \SI{6.67}{\micro\second})$.
The time-difference-of-arrival (TDOA) between two transceivers $i$ and $j$ is modelled as:
\begin{align}\label{eq:tdoa_model}
    \delta_{ij} = t_i - t_j = -\frac{1}{c}\myFrameVec{d}{}{\mathrm{T}}\left(\myFrameVec{p}{HYDRO,i}{b}-\myFrameVec{p}{HYDRO,j}{b}\right), 
\end{align}
with $\myFrameVec{d}{}{}$ as the direction vector from nanoAUV's $i$'th transceiver and the (home) USBL transponder. The USBL measures TDOAs between all $N=5$ transceivers:
\begin{align}
\myFrameVecTilde{\Delta}{t}{} = [
\myFrameScalarTilde{\delta}{12}{},~\myFrameScalarTilde{\delta}{13}{},~\text{...},~\myFrameScalarTilde{\delta}{N-1,N}{}]^\mathrm{T} + \myFrameVec{w}{\Delta_t}{},
\end{align}
with $\myFrameVec{w}{\Delta_t}{}$ as the TDOA measurement noise ($1\sigma = \SI{0.3}{\micro\second})$.

Given the measured TDOA vector $\myFrameVecTilde{\Delta}{t}{}$ and the matrix-form $\myFrameVec{S}{}{} = \myFrameVec{D}{}{}\cdot\myFrameVec{P}{HYDRO}{}$ of \eqref{eq:tdoa_model} for five transceivers,
the direction vector can be calculated via least-squares:
\begin{align}
\myFrameVec{d}{}{} = -c \cdot\myFrameVec{S}{}{+}\cdot\myFrameVecTilde{\Delta}{t}{} \quad \text{with: }\myFrameVec{S}{}{+} = (\myFrameVec{S}{}{\mathrm{T}}\cdot\myFrameVec{S}{}{})^{-1}\cdot\myFrameVec{S}{}{\mathrm{T}}
\end{align}
and hence the simulated position measurement of the nanoAUV transceiver modem:
\begin{align}\label{eq:usbl_simulation_model}
    \myFrameVecTilde{\eta}{USBL,AUV}{} = \frac{c}{2} \cdot \myFrameScalarTilde{RTT}{}{} \cdot \frac{\myFrameVec{d}{}{}}{\|\myFrameVec{d}{}{}\|}.
\end{align}
For standard USBL operations, the transceiver is installed at the surface/operator site. This induces additional motions which need to be compensated. For the nanoAUV under-ice operations, the transceiver is installed on the nanoAUV. Furthermore, the (home) transponder is assumed to remain fixed in the ice hole. Hence, rotational motions are assumed to be nearly zero and therefore do not need to be compensated.

\section{Path Planning, Guidance, and Control}\label{sec:path_planning_guidance_and_control}

\begin{figure*}[htb]
    \centering
    \includegraphics[width=0.95\textwidth]{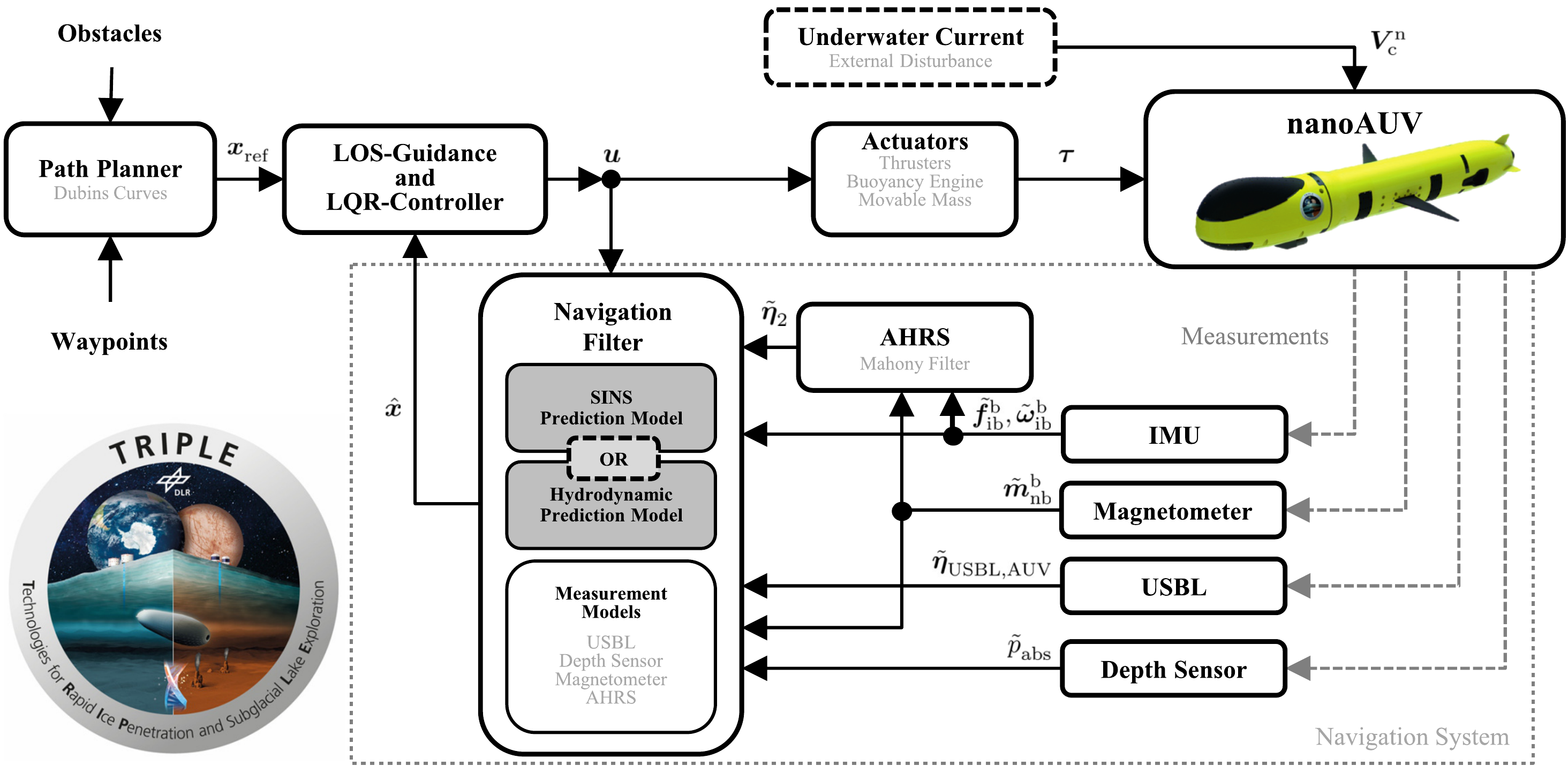}
    \caption{Proposed GNC system for the TRIPLE-nanoAUV (Credits nanoAUV hull design: Sebastian Meckel, MARUM).}
    \label{fig:navigation_filter}
\end{figure*}
The preliminary path planning, guidance, and control framework is designed with low computational effort and robustness as the primary goals according to the mission requirements. It is discussed more in \cite{Stenger.2022}.  

\subsection{Path Planning}
We use Dubins curves to find the shortest path denoted by $\myFrameVec{x}{ref}{}$ between the current estimated vehicle state and the next waypoint in the horizontal plane. The depth reference is generated using linear interpolation between starting and goal depth (cf.,e.g., \cite{Valavanis.2015}). 
The minimum turning radius within the Dubins path planner ensures that the AUV can track the planned path. At this point, a constant surge velocity is planned.    

\subsection{Guidance}
The guidance module resolves the underactuation of the nanoAUV. Line of sight guidance with adaptive sideslip compensation according to \cite{FossenThorI..2015} is used to calculate the yaw angle reference $\myFrameScalar{\psi}{d}{}$ for the lower level controller. Smooth and accurate estimation of the velocities $v$ and $u$ is one of the key tasks of the navigation filter:
\begin{align} \label{eq:LOS_yaw}
    \myFrameScalar{\psi}{d}{} =  \myFrameScalar{\gamma}{p}{} + \mathrm{atan}(-\frac{\myFrameScalar{h}{e}{}}{\Delta}) -\myFrameScalar{\beta}{est}{} \text{.}
\end{align}
The first term in \eqref{eq:LOS_yaw} represents the path angle $\myFrameScalar{\gamma}{p}{}$ and the second term aims at reducing the horizontal cross-track error $\myFrameScalar{h}{e}{}$ (i.e., the shortest distance between the current AUV position and the reference path on the horizontal plane), with $\Delta$ being a tuning parameter. The side slip angle $\myFrameScalar{\beta}{k}{} = \mathrm{atan}(\frac{v}{u})$ occurs due to current induced drift. 
The smooth and accurate estimation of the velocities $v$ and $u$ is one of the key tasks of the navigation filter (cf. Sec. \ref{sec:navigation_filter}). 
The vertical guidance is treated in an analog manner.     

\subsection{Control}
The well-known LQR method is used to synthesize a state feedback controller to track the surge velocity provided by the planner and the pitch and yaw angles provided by the guidance modules. The controlled states are augmented with integral error states to achieve stationary accuracy in the presence of HMM mismatch. The model for LQR synthesis is obtained by discretizing and linearizing \eqref{eq:fossen_model}. The LQR controller only yields inputs for the thrusters. The movable mass, as well as the buoyancy engine, are controlled in a strictly feed-forward manner. 

\section{Navigation Filter}\label{sec:navigation_filter}
The navigation filter is implemented as a UKF as in \cite{fanelli2019} and provides state estimates to the guidance and control system shown in Fig.\,\ref{fig:navigation_filter} and discussed in more detail in \cite{Stenger.2022}.  
We implement two filters: a UKF with a prediction model based on SINS and a UKF with an HMM (Fossen) as a prediction model. The latter approach does not fuse MAG measurements within the correction step but the attitude solution of an external attitude heading reference system (AHRS). This is motivated by the fact that accelerometer and gyroscope measurements are typically integrated within the prediction step of a SINS, which is not applicable to the HMM prediction. Fusing accelerometer, gyroscope, and MAG measurements within the AHRS is much more computationally efficient compared to three additional state updates within the UKF with HMM prediction. This decoupled filter architecture is inspired by \cite{fanelli2019}. The AHRS is implemented as a Mahony filter which is also found in \cite{fanelli2019}. 

For both filter architectures, quaternions are used to represent the attitude which gives $\myFrameVec{\eta}{2}{} = [\eta, \varepsilon_1, \varepsilon_2, \varepsilon_3]^\mathrm{T}$. This avoids potential gimbal locks and gives higher accuracy according to \cite[p.\,324]{titterton2004}. A UKF requires special treatment of quaternion operations. They are handled as by \cite{cheon2007}. Both process models are discretized with the explicit Euler method. 
Except for the measurement update with MAG and AHRS, both filters implement the same measurement models for the pressure sensor and USBL.

\subsubsection{SINS Prediction Model}
The process model for the first UKF is based on SINS as found in \cite{titterton2004}. Measured specific force $\myFrameVecTilde{f}{ib}{b}$ and rotation rate $\myFrameVecTilde{\omega}{ib}{b}$ are inputs $\myFrameVec{u}{}{}$ for the process model. The accelerometer $\myFrameVec{b}{a}{}$ and gyroscope bias $\myFrameVec{b}{g}{}$ are modeled as random-walk. The rotation rates $\myFrameVec{\nu}{2}{}$ are not part of the SINS prediction model but are constructed by subtracting the gyroscope bias from the measurement: $\myFrameVec{\nu}{2}{} = \myFrameVecTilde{\omega}{ib}{b}-\myFrameVec{b}{g}{}$.
The process model for the SINS-based filter is given by:
\begin{align}\label{eq:strapdown}
	\underbrace{
		\pVecFive
		{\myFrameVecDot{\nu}{1}{}}
		{\myFrameVecDot{\eta}{1}{}}
		{\myFrameVecDot{q}{b}{n}}
		{\myFrameVecDot{b}{a}{}}
		{\myFrameVecDot{b}{g}{}}
	}_{\myFrameVecDot{x}{}{}}
	=&
	\underbrace{
		\pVecFive
		{(\myFrameVecTilde{f}{ib}{b}-\myFrameVec{b}{a}{})-\myFrameVec{C}{n}{b}(\myFrameVec{q}{b}{n})\myFrameVec{g}{nb}{n}}
		{\myFrameVec{C}{b}{n}(\myFrameVec{q}{b}{n})\myFrameVec{\nu}{1}{}}
		{ \frac{1}{2}\myFrameVec{\Omega}{}{}\left(0,\left[\myFrameVec{\omega}{ib}{b} - \myFrameVec{b}{g}{}\right]\right)\cdot\myFrameVec{q}{b}{n}}
		{\myFrameVec{0}{}{}}
		{\myFrameVec{0}{}{}}
	}_{\myFrameVec{f}{}{}(\myFrameVec{x}{}{},\myFrameVec{u}{}{})} +
	\myFrameVec{G}{}{}
	\underbrace{
		\pVecFour
		{\myFrameVec{w}{a}{}}
		{\myFrameVec{w}{g}{}}
		{\myFrameVec{w}{b_a}{}}
		{\myFrameVec{w}{b_{g}}{}}
	}_{\myFrameVec{w}{IMU}{}},
\end{align}
with rotation matrix from body to NED frame $\myFrameVec{C}{b}{n}(\myFrameVec{q}{b}{n})$ and skew operator $\myFrameVec{\Omega}{}{}$. Coriolis and transport rate are neglected for this work. IMU noise $\myFrameVec{w}{IMU}{}$ is mapped by $\myFrameVec{G}{}{}$ to the states.

\subsubsection{Hydrodynamic Motion Prediction Model}
The process model for the second UKF is based on the HMM from \eqref{eq:fossen_model}. Generalized forces and moments $\myFrameVec{\tau}{}{}$ are inputs $\myFrameVec{u}{}{}$ for the process model. Water current surge $\myFrameScalar{u}{c}{n}$ and sway $\myFrameScalar{v}{c}{n}$ components are modeled by a random walk. The heave component $\myFrameScalar{w}{c}{n}$ is neglected. The process model for the HMM-based filter is given by:
\begin{align}\label{eq:fossen}
	\underbrace{
		\pVecSix
	    {\myFrameVecDot{\nu}{1}{}}
		{\myFrameVecDot{\nu}{2}{}}
		{\myFrameVecDot{\eta}{1}{}}
		{\myFrameVecDot{q}{b}{n}}
		{\myFrameScalarDot{u}{c}{n}}
		{\myFrameScalarDot{v}{c}{n}}
	}_{\myFrameVecDot{x}{}{}}
	=&
	\underbrace{
		\pVecSix
		{\text{Fossen from \eqref{eq:fossen_model}}}
		{\text{Fossen from \eqref{eq:fossen_model}}}
		{\myFrameVec{C}{b}{n}(\myFrameVec{q}{b}{n})}
		{\myFrameVec{T}{q}{}(\myFrameVec{q}{b}{n})}
		{0}
		{0}
	}_{\myFrameVec{f}{}{}(\myFrameVec{x}{}{},\myFrameVec{u}{}{})} + 
	\myFrameVec{G}{}{}
	\underbrace{
		\pVecSix
	    {\myFrameVec{w}{\nu_{1}}{}}
		{\myFrameVec{w}{\nu_2}{}}
		{\myFrameVec{w}{\eta_{1}}{}}
		{\myFrameVec{w}{\eta_2}{}}
		{\myFrameScalar{w}{u_c}{}}
		{\myFrameScalar{w}{v_c}{}}
	}_{\myFrameVec{w}{HYDRO}{}},
\end{align}
with $\myFrameVec{T}{q}{}$ as angular velocity transformation matrix according to \cite[p.\,29]{fossen2011} and $\myFrameVec{w}{HYDRO}{}$ as the process noise of the HMM.

\subsubsection{Depth Sensor Measurement Model}
The estimated depth sensor measurement is:
\begin{align}
    \myFrameScalarHat{p}{abs}{} = \myFrameScalar{k}{p}{}\cdot \myFrameScalarHat{d}{}{} + \myFrameScalar{p}{0}{} + \myFrameScalar{v}{press}{},
\end{align}
with $\myFrameScalar{v}{press}{}$ the white noise of the pressure sensor.

\subsubsection{MAG Measurement Model}
The MAG measurements are inputs to the filter with the SINS-based prediction model.
The estimated MAG measurement is calculated by:
\begin{align}\label{eq:magnetometer_meas_model}
    \myFrameVecHat{m}{nb}{b} = \myFrameVec{C}{n}{b}(\myFrameVecHat{q}{b}{n})\cdot \myFrameVec{m}{nb}{n}+ \myFrameVec{v}{MAG}{},
\end{align}
with MAG noise $\myFrameVec{v}{MAG}{}$.

\subsubsection{AHRS Measurement Model}
The Mahony filter outputs an estimated quaternion $\myFrameVecHat{q}{b,AHRS}{n}$ which is transformed to Euler angles $\myFrameVecHat{\eta}{2}{}$.
These are the external AHRS inputs to the filter with HMM prediction. The resulting measurement model is:
\begin{align}
     \myFrameVecHat{\eta}{2,AHRS}{} = \mathrm{Euler}(\myFrameVecHat{q}{b}{n}) + \myFrameVec{v}{AHRS}{},
\end{align}
with $\myFrameVec{v}{AHRS}{}$ as the measurement noise of the AHRS. The "$\mathrm{Euler}$" operation converts the quaternion to Euler angles according to, i.e., \cite[p.\,33-34]{fossen2011}. 
BO-tuned filter parameters are in Tab.\,\ref{tab:mahony_params}. 
Around $\pm\pi$ measured heading $\myFrameScalarTilde{\psi}{}{}$ and estimated heading $\myFrameScalarHat{\psi}{}{}$ can 
have different signs. A correction method for the Kalman innovation term is discussed in \cite{nitsch2021b}.

\subsubsection{USBL Measurement Model}
The estimated USBL measurement is given by:
\begin{align}\label{eq:usbl_simulation}
    \myFrameVecHat{\eta}{USBL,AUV}{} = \myFrameVecHat{\eta}{1}{} + \myFrameVec{v}{USBL}{},
\end{align}
with USBL noise $\myFrameVec{v}{USBL}{}$. Lever arms are neglected.

\subsection{Nominal Filter Parametrization}
For the SINS filter, the UKF parameter $\alpha$ is set to \SI{1}{} and for the HMM filter to \SI{1e-3}{}. These values have proven to be the most stable for the respective filters. The remaining parameters are set according to \cite{nitsch2021b}. Values for the initial covariance matrices $\myFrameVec{P}{0}{}$ are set according to Tab.\,\ref{tab:initial_noise_values}.
The values for the process noise matrix $\myFrameVec{Q}{IMU}{}$ in Tab.\,\ref{tab:process_noise_values} base on the Allan variance analysis as shown in Sec.\,\ref{subsec:sensor_simulation_models}. The values for the process noise matrix $\myFrameVec{Q}{HYDRO}{}$ are a pure designer's choice.
\begin{table}[!ht]
\centering
\begin{minipage}[t]{0.48\linewidth}\centering
\caption{Process model noise}
\label{tab:process_noise_values}
\begin{tabular}{ l l }
\toprule
\textbf{Noise}  &  1$\sigma$ STD \\
\midrule
$\myFrameVec{w}{b_{a|g}}{}$       & Allan Var. from Fig.\,\ref{fig:allan_variance} \\
$\myFrameVec{w}{a|g}{}$       & Allan Var. from Fig.\,\ref{fig:allan_variance}             \\
\midrule
$\myFrameVec{w}{\nu_1}{}$   & \SI{1e-3}{\radian\per\second}              \\
$\myFrameVec{w}{\nu_2}{}$   & \SI{1e-3}{\metre\per\second}           \\
$\myFrameVec{w}{\eta_1}{}$  & \SI{0.01}{\metre}               \\
$\myFrameVec{w}{\eta_2}{}$  & \SI{0.01}{\degree}               \\
$\myFrameScalar{w}{u_c|v_c}{}$     & \SI{1e-3}{\metre\per\second} \\
\bottomrule
\end{tabular}
\end{minipage}\hfill
\begin{minipage}[t]{0.48\linewidth}\centering
\caption{Initial standard deviations}
\label{tab:initial_noise_values}
\begin{tabular}{ l l }
\toprule
\textbf{State}    &  1$\sigma$ STD \\ 
\midrule
$\myFrameVec{\nu}{1}{}$ & \SI{0.25}{\metre\per\second}\\
$\myFrameVec{\nu}{2}{}$ & \SI{0.5}{\radian\per\second}\\
$\myFrameVec{\eta}{1}{}$ & \SI{0.5}{\metre}\\
$\myFrameVec{\eta}{2}{}$ & \SI{2.5}{\degree}\\
$\myFrameScalar{u}{c}{}$ & \SI{5}{\metre\per\second}\\
$\myFrameScalar{v}{c}{}$ & \SI{5}{\metre\per\second}\\
$\myFrameVec{b}{a}{}$ & \SI{0.1}{\metre\per\second\squared}\\
$\myFrameVec{b}{g}{}$ & \SI{1e-3}{\degree\per\second}\\
\bottomrule
\end{tabular}
\end{minipage}
\end{table}
The values of the matrices $\myFrameVec{R}{}{}$ for the measurement models are found in Tab.\,\ref{tab:measurement_noise_values}. The USBL covariance matrix $\myFrameVec{R}{USBL}{}$ is weighted depending on the distance. The EvoLogics USBL delivers a positional accuracy which we mimic by propagating the RTT and TDOA noise vectors through the nonlinear model in \eqref{eq:usbl_simulation_model}. This accuracy value is multiplied with $\myFrameVec{R}{USBL}{}$. The matrices $\myFrameScalar{R}{press}{}$,  $\myFrameVec{R}{MAG}{}$ and $\myFrameVec{R}{AHRS}{}$ are static.
\begin{table}[!ht]
\centering
\begin{minipage}[t]{0.48\linewidth}\centering
\caption{Measurement model noise}
\label{tab:measurement_noise_values}
\begin{tabular}{ l l}
\toprule
\textbf{Noise}  &  1$\sigma$ STD \\
\midrule
			$\myFrameVec{v}{USBL}{}$ & \SI{1}{\metre} \\
			$\myFrameScalar{v}{press}{}$ & \SI{2500}{\pascal} \\
			$\myFrameVec{v}{MAG}{}$ & Allan Var. from Fig.\,\ref{fig:allan_variance}\\
			$\myFrameVec{v}{AHRS}{}$ & $[0.25, 0.25, 0.80]$ \SI{}{\degree}\\
\bottomrule
\end{tabular}
\end{minipage}\hfill
\begin{minipage}[t]{0.48\linewidth}\centering
\caption{Mahony filter parameters}
\label{tab:mahony_params}
\begin{tabular}{ l l }
\toprule
\textbf{Parameter} & Value \\ 
\midrule
			$k_P$ & 55.7037 \\
			$k_I$ & 48.3934 \\
			$k_1$ & 0.4828 \\
			$k_2$ & 0.0749 \\
\bottomrule
\end{tabular}
\end{minipage}
\end{table}

\subsection{Parameter Tuning with Bayesian Optimization}
The performance of both filter versions relies on the values for the tuning parameters $\myFrameVec{a}{}{}$. We optimize five tuning parameters $a_1 \dots a_5$ for the matrices
$\myFrameVec{Q}{\nu_1}{}$, $\myFrameVec{Q}{\eta_1}{}$, $\myFrameVec{Q}{\nu_2}{}$, $\myFrameVec{Q}{\eta_2}{}$, $\myFrameVec{R}{AHRS}{}$ of the HMM filter and the matrices
$\myFrameVec{Q}{a}{}$, $\myFrameVec{Q}{g}{}$, $\myFrameVec{Q}{b_a}{}$, $\myFrameVec{Q}{b_g}{}$, $\myFrameVec{R}{MAG}{}$ of the SINS filter.

To find the optimal parameters $\myFrameVec{a}{}{\star}$, the tuning problem is formulated as follows:
\begin{align} \label{eq:generalOptProb}
  \quad \myFrameVec{a}{}{\star} =  \mathrm{argmin} \quad \quad \quad &J(\myFrameVec{a}{}{}) \\
\text{s.t.} \, \, \, \, \, \, \, \quad \quad   \myFrameVec{a}{\mathrm{min}}{} &\leq \myFrameVec{a}{}{} \leq \myFrameVec{a}{\mathrm{max}}{} \nonumber  \\ 
  g(\myFrameVec{a}{}{}) &\leq g_{\mathrm{max}} \nonumber \\ 
  l(\myFrameVec{a}{}{}) &= 1 \nonumber .
\end{align}
For the open-loop case, $J(\myFrameVec{a}{}{})$ is the RMS error between estimated $\myFrameVecHat{\eta}{1}{}$ and ground truth $\myFrameVec{\eta}{1}{}$ positions. In addition to box constraints, a constraint on the maximum RMS Euler angle error $g(\myFrameVec{a}{}{}) \leq g_{\mathrm{max}}=\SI{3}{\degree}$ is added for the open-loop case. 
For the closed-loop case, the objective function $J(\myFrameVec{a}{}{})$ is the maximum deviation from the reference path (tracking error). It is evaluated using the simulation environment described in Sec.\,\ref{sec:simulation_environment}. Furthermore, simulations may be aborted due to poor parametrizations resulting in the nanoAUV not reaching its goal state or the filter diverging. In this case, objective function value and constraints are unavailable, which poses an additional challenge. This type of constraint is known as a \textit{crash constraint} and denoted by $l(\myFrameVec{a}{}{}) = 1$. 

The optimization problem in \eqref{eq:generalOptProb} is solved using BO with constrained max-value entropy search according to \cite{Perrone.15.10.2109}, as the acquisition function and a squared-exponential kernel with automated relevance detection for the GP-model. Crash constraints are handled using virtual data points. The details of the BO method are explained in \cite{Stenger.2022}. For the unconstrained case, details can be found in \cite{Stenger.2022b}.    
The sampling efficiency of BO is also compared with that of constrained PSO\footnote{Toolbox:\,\,\,\url{https://de.mathworks.com/matlabcentral/fileexchange/25986-constrained-particle-swarm-optimization}}.

\section{Results}\label{sec:results}
\begin{figure}[ht]
    \centering
    \includegraphics[width=0.8\columnwidth]{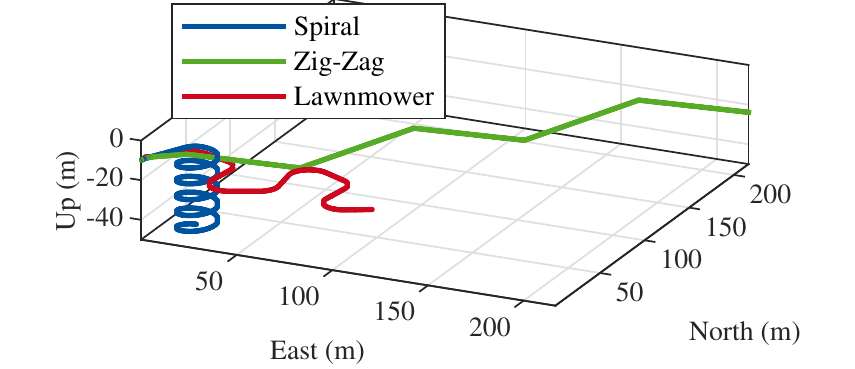}
    \caption{Trajectories for optimization and Monte Carlo simulations.}
    \label{fig:trajectories}
\end{figure}

We first show the optimization results for the solution of \eqref{eq:generalOptProb} for both navigation filters and the open-loop and closed-loop cases. Within the cost function, three simulations are performed for spiral, zig-zag, and lawnmower trajectories, as shown in Fig.\,\ref{fig:trajectories}. For each trajectory, a new seed is chosen for sensor noise, underwater current, and model mismatch of the nominal HMM. A USBL outage of \SI{30}{\second} is generated. Estimated states $\myFrameVecHat{\nu}{1|2}{}$ and $\myFrameVecHat{\eta}{1|2}{}$ are initialized with ground-truth. The states $\myFrameVecHat{b}{a|g}{}$, $\myFrameScalarHat{u}{c}{}$ and $\myFrameScalarHat{v}{c}{}$ are initialized with zero. All trajectories start at $\myFrameVec{\eta}{1}{} = [0,0,10]^\mathrm{T}$.
With the optimized parameter sets, extensive Monte Carlo simulations with two USBL outages of \SI{10}{\second} and \SI{30}{\second} are performed on the same trajectories (\ref{sec:resMonteCarlo}).

\subsection{Filter Parameter Optimization}
\begin{figure}[ht]
    \centering
    \includegraphics[width=1\columnwidth]{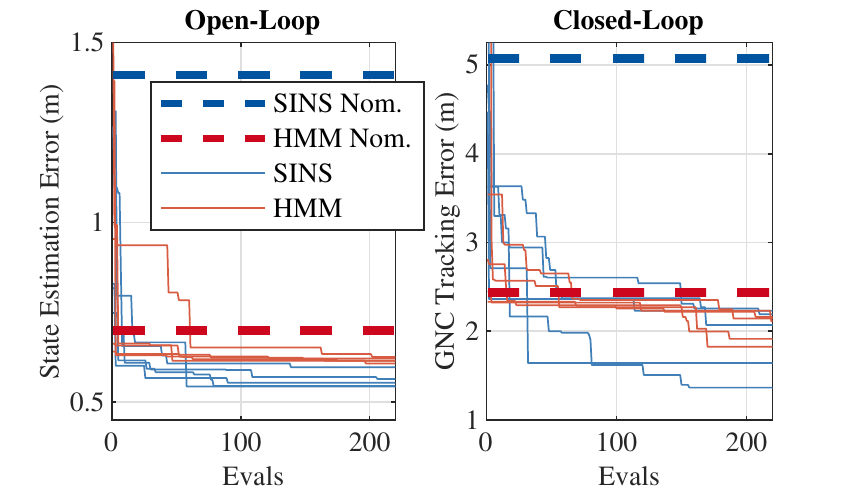}
    \caption{BO tuning results for open- and closed-loop performance. The dashed lines represent the errors for nominal parametrization.}
    \label{fig:opt_results}
\end{figure}

\begin{figure}[ht]
    \centering
    \includegraphics[width=1\columnwidth]{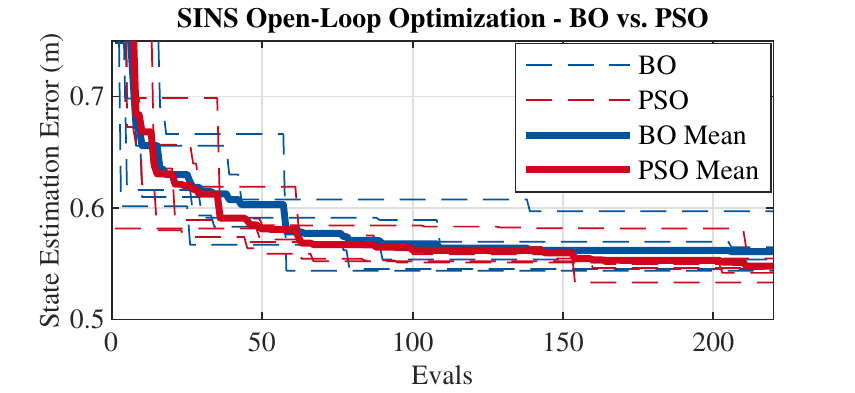}
    \caption{Comparison of BO and PSO for SINS open-loop optimization.}
    \label{fig:opt_results_BOPSO}
\end{figure}

The results of the BO parameter tuning are shown in Tab.\,\ref{tab:opt_results} and Fig.\,\ref{fig:opt_results}. We perform five optimization runs, each with 225 evaluations of $J(\myFrameVec{a}{}{})$ for both filters. An evaluation takes about \SI{200}{\second}. Results indicate that the nominal parametrization for the SINS-based filter is worse than the optimum. Especially the tracking error (RMS) benefits from BO parameter tuning. The HMM filter benefits only slightly from parameter tuning for the state estimation accuracy (RMS) in open-loop but also shows a significant improvement in tracking performance for the closed-loop.
\vspace{-0.05cm}
Fig.\,\ref{fig:opt_results} indicates that the nominal parametrization for the HMM filter has already been chosen quite well by the designer for the assumed model mismatch. This is not the case for the SINS filter, although the parametrization is based on Allan variance analysis of the IMU noise. For open-loop optimization, the state estimation error of both filters is in close range. For the closed-loop case, the SINS filter achieves \SI{46}{\centi\metre} better tracking performance. However, the optimum found for the closed-loop case scatters in a range between \SI{1.366}{\metre} and \SI{2.231}{\metre}. Fig.\,\ref{fig:opt_results_BOPSO} indicates that with 225 evaluations, the global optimum is not found consistently with BO and PSO. In contrast to \cite{Stenger.2022b}, PSO slightly outperforms BO on average. However, differences are not statistically significant, indicating that \textbf{BO and PSO may be equally suited for the presented tuning problem}. 

\begin{table}[!ht]
\caption{BO tuning results for $J(\myFrameVec{a}{}{\star})$ - Best (worst) of five optimization runs}
\label{tab:opt_results}
\centering
\begin{tabular}{ l c c}
\toprule
\textbf{Filter Case}  &  State Estimation Error & GNC Tracking Error \\
\midrule
SINS Nom.       & \SI{1.410}{\metre}\,(\SI{1.410}{\metre})  & \SI{5.072}{\metre}\,(\SI{5.072}{\metre}) \\
SINS Opt.       & \SI{0.544}{\metre}\,(\SI{0.597}{\metre})  &   \SI{1.366}{\metre}\,(\SI{2.231}{\metre}) \\
HMM Nom.     & \SI{0.700}{\metre\,(\SI{0.700}{\metre})}  &   \SI{2.439}{\metre}\,(\SI{2.439}{\metre})\\
HMM Opt.     & \SI{0.608}{\metre}\,(\SI{0.626}{\metre})  &   \SI{1.826}{\metre}\,(\SI{2.162}{\metre})\\
\bottomrule
\end{tabular}
\end{table}

\subsection{Monte Carlo Performance Evaluations} \label{sec:resMonteCarlo}
\begin{figure}[ht]
    \centering
    \includegraphics[width=0.9\columnwidth]{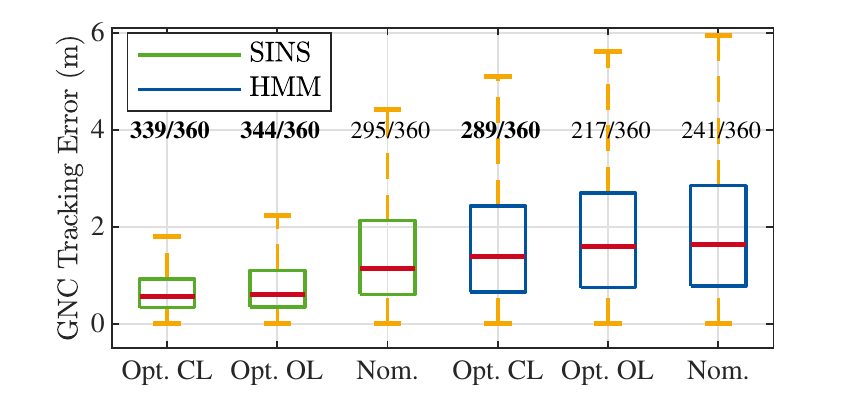}
    \caption{Box plot of closed-loop tracking errors and reached goals from Monte Carlo runs. The CL-optimized filters perform significantly better.}
    \label{fig:monte_carlo_cl}
\end{figure}

Monte Carlo simulations are performed with all three trajectories for all filter configurations. Each simulation is executed for a different seed, resulting in 360 various Monte Carlo runs. From GNC's point of view, there are two targets to be achieved: reaching the final goal state and minimizing the tracking error. In Fig.\,\ref{fig:monte_carlo_cl}, the tracking errors are shown for the Monte Carlo runs \textbf{that reach the final goal}, which is given by the path planning algorithm. Some optimized parametrizations lead to an unstable closed-loop, which results in the goal not being reached. The tracking error calculation is not performed for these parametrizations since it increases far above reasonable values.
These validation results indicate that for BO-tuned parameterizations, the number of goals reached can be increased and the tracking error minimized. The goal is considered to be reached when the nanoAUV is within a radius of $\SI{10}{\metre}$ around the final waypoint. This distance is currently assumed to be within range of the proposed docking system. For the SINS filter, a comparable number of goals is reached either for open- or closed-loop tuned parameters (339 and 344/360). Compared to nominal SINS parametrization, the number of reached goals is drastically increased (only 295/360). For the HMM filter, closed-loop parameter tuning shows the best results concerning reached goals and tracking performance. Open-loop tuning, in this case, even leads to significantly fewer reached goals than with nominal parameters.
This \textbf{confirms our hypothesis} that a filter optimized for state estimation accuracy does not necessarily achieves good tracking performance in closed-loop operation.

For the proposed simulation scenario, we conclude that the SINS filter performs better than the HMM filter in reaching the final state and tracking performance. However, we cannot state that SINS prediction is more beneficial than HMM prediction because the HMM performance in our simulations strongly depends on the generated model mismatch.
However, both prediction approaches benefit from BO-based parameter tuning for the open- and closed-loop cases.
\vspace{-0.05cm}

\section{Conclusions}\label{sec:conclusions}
In this contribution, we have demonstrated BO-based automated parameter tuning of SINS- and HMM-based navigation filters for an underactuated miniature AUV (nanoAUV).
We tuned the filters for state estimation accuracy in the open-loop and tracking performance in the closed-loop case. The results show that automated tuning leads to significantly better GNC tracking results. For this application, PSO was shown to be equally suited as BO to solve the tuning problem. The nanoAUV can reach the target state in more cases under different water current profiles and USBL outages. Moreover, it confirms the hypothesis that "naive" tuning for optimal state estimation accuracy does not always lead to optimal tracking accuracy once the GNC loop is closed. Hence the state estimate of the filter is used as feedback.

Future work will focus on combining SINS, and HMM, i.e., like in \cite{hegrenaes2008}. The HMM can limit the quadratically increasing position errors of the SINS, especially if the USBL fails.
We plan experiments with a nanoAUV prototype equipped with the proposed sensor suite. The AUV will be tracked with an underwater camera system to generate ground truth data and to identify an HMM model. The proposed BO-automated tuning method will be used to find an optimized parameter set for the navigation filter.
We are interested in determining how position, velocity, and attitude estimation errors relate to the GNC tracking error. We plan to find the Pareto front between these metrics and extract tuning heuristics for filter designers to increase the GNC tracking performance. Furthermore, multi-objective tuning of state estimation errors beyond the position error will be investigated.

\bibliography{ifacconf} 

\end{document}